# Direct Digital Design of Loop-Shaping Filters for Sampled Control Systems

Hugh L. Kennedy*

*Abstract*—A controller design technique for shaping the frequency response of a process is described. A general linear model (GLM) is used to define the form of a lag or lead compensator in discrete time using a prescribed set of basis functions. The model is then transformed via the complex *z*-domain into a difference equation for a recursive digital filter with an infinite impulse response (IIR). A polynomial basis set is better for shaping the frequency response in the near-zero region; whereas a sinusoidal basis set is better for defining the response at arbitrary frequencies. The proposed compensator design method is more flexible than existing low-order approaches and more suitable than other general-purpose high-order methods. Performance of the resulting controller is compared with digital proportional-integral-differential (PID) and linear-state-space (LSS) algorithms in a real motor-control application.

1. **Introduction**

Frequency-domain controller-design provides a direct link between system measurement and system analysis processes; it also allows feedback controller synthesis to be viewed as a filter design problem [1]-[9]. While the transfer function of the process to be controlled (i.e. the plant) cannot be changed, the transfer function of the compensator can be contrived to shape the open-loop frequency-response to yield an integrated closed-loop system with the desired transient/steady-state response and satisfactory stability (i.e. gain and phase) margins. This perspective has the potential to be particularly useful when designing digital compensators for computer implementation in discrete-time control systems because it provides a direct link to digital signal processing (DSP) techniques.

However, digital compensators often have unique and subtle requirements in the time and frequency domains that are not properly considered in exiting general-purpose digital-filter design methods [10]. For instance, phase lead is more important than phase linearity in a digital compensator; as a consequence, discontinuities and rapid changes in the frequency response are avoided (e.g. closely-spaced zeros on the unit circle and narrow transition bands) to ensure that phase and magnitude requirements are jointly satisfied.

Furthermore, plant transfer functions are rarely known precisely thus industrial controller design is an iterative and ongoing process. The ability to quickly fine-tune compensators in deployed control systems, using intuitive interfaces and dependable software is desirable, which is one of the reasons why digital proportional-integral-differential (PID) control is so popular.

Frequency-domain compensator design is guided by the following basic principles [11]:

- High open-loop gain is desirable wherever the closed-loop controller is required to exert its influence; for instance:

---

* The corresponding author is with the Defence and Systems Institute, School of Engineering, University of South Australia, Mawson Lakes, Australia (e-mail: hugh.kennedy@unisa.edu.au; fax: +61 8 8302 5344).

- At low-frequencies, to reduce steady-state errors when tracking simple reference-inputs (e.g. steps and ramps) in servomechanisms;
- At low-to-medium frequencies, for disturbance attenuation/rejection in regulators and an improved transient response in servomechanisms (i.e. faster rise time, faster settling time and reduced overshoot).
- Low open-loop gain at high frequencies is required to reduce sensor measurement noise and interference.
- The open-loop transfer function must attenuate frequencies wherever the phase lag exceeds 180 degrees, to guarantee stability in a closed loop configuration. This ensures that stable negative-feedback does not become unstable positive-feedback. The margins of stability, thus the ability of the closed-loop system to remain stable in the presence of plant modelling errors (e.g. unexpected system gains and delays), increase as the attenuation increases.

Lag and lead compensators are commonly used as loop-shaping instruments – Lag compensators are low-pass filters, with a backward phase shift over the transition band; lead compensators are high-pass filters, with a forward phase-shift over the transition band. Lag compensators are used to attenuate high-frequencies, with a phase lag as an undesirable side-effect; whereas lead filters are used to provide a forward phase-shift, with high-frequency amplification as an undesirable side-effect.

Lag compensators behave somewhat like an integrator; whereas lead compensators have differentiator-like properties. They may therefore be combined to emulate the properties of PID-type controllers. PID controllers are indeed a specialized compensator form, with a marginally stable pole at $z = 1$ due to the integrator and up to two zeros; however in principle, general lag/lead compensators may involve the placement of any number of poles and zeros to achieve the desired frequency response. The absence of a pure integrator in a lag/lead compensator made it a popular type of analogue controller. As charge cannot be accumulated forever without dissipation, active circuitry is required to implement an analogue integrator. While this constraint does not directly apply to digital systems, care must still be taken to avoid numerical overflow due to persistent plant-output/reference-input offsets and integrator wind-up due to actuator saturation [12].

Lag and lead components are clearly complementary. It would be ideal to have single filter cut high-frequency gain (for improved noise immunity) *and* provide a phase lead (for improved stability) to permit the application of a large controller gain (for improved transient and stead-state behaviour); however in practice, a compromise must be reached with lead and/or lag filters tuned to yield an appropriate balance.

Simple first-order compensators are designed through the careful placement of a real pole and a real zero in either the complex *s* or complex *z* domains. Expressions relating the zero/pole locations to the natural frequency, damping ratio and phase lead/lag are available to aid this process in the *s* domain [2], [3]. First-order compensators of the same type may be cascaded to intensify their effect; or a heterogeneous network may be used for a more moderate effect. It has been shown that second-order lead/lag filters with complex poles and zeros provides greater design flexibility [2], [4]. Graphical methods for the design of lead-lag compensators in the *s* and *z* domains have also been proposed [5], [6]. The classical design methods described so far are restricted to first- or second-order filters; furthermore, there are very few options for direct design in the *z* domain.

A simple and flexible process for designing low-order lead and lag compensators, directly in the digital domain, is described in this paper. The compensators are designed using polynomial basis functions or sinusoidal basis functions in the time domain. Polynomials are good for manipulating the low-frequency response, whereas sinusoids are better for arbitrary frequencies. The same general design approach is used in both cases – Recursive filters are realized by fitting the functions to the input signal, in a weighted linear-least-squares sense. The fitted functions are then evaluated at some point in the future (for a phase lead) or the past (for a phase lag). The proposed design approach simply involves some *z* polynomial manipulation and the inverse of a small symmetric matrix; it does not require numerical optimization followed by discretization, unlike classical $\mathcal{H}_\infty$ loop-shaping approaches [7], [8].

In the next section (Section 2), the proposed filter-design method is summarized in the context of relevant prior work. Details of the filter design process are given in Section 3 – Subsection 3.1 for polynomials and Subsection 3.2 for sinusoids. The derivation begins with the FIR case to illustrate the basic concept; the weighting function is then introduced and the desired IIR filters are derived via the $z$ domain. Some general tuning considerations are also discussed in Subsection 3.3. Simulated examples are then used in Section 4 to illustrate the main design and tuning principles; while an electric motor is used in Section 5 to highlight some practical considerations in a real control application. Performance of the proposed design approach in the motor control problem is compared with some candidate design alternatives of similar complexity; namely, an empirically-tuned PID controller and a linear-state-space (LSS) servomechanism [13]. The paper closes with a summary, some recommendations and concluding remarks in Section 6.

## 2. Overview of the Proposed Approach

Many different digital filters have been derived using polynomial interpolation, in one form or another (e.g. linear least-squares regression, Taylor series and Lagrange interpolation) – from integrators and differentiators through to low-pass and high-pass filters, to name just a few [14]-[19]. Predictive polynomial filters have been proposed to compensate for delays in networked control systems [20]; and to have a low-pass filtering effect with minimal group delay or an all-pass effect with minimal gain distortion in measurement instruments [19], [21]; however, they have not been directly applied as lag/lead compensator components in a closed-loop control context. Predictive polynomial filters with a finite impulse response are generally favoured [19]; in [22], the FIR filters are implemented recursively using a real pole on the unit circle, cancelled by a zero; in [23] an efficient polyphase approach is described.

Predictive filters are derived using divided differences in [19], so that the interpolating polynomial passes through each sample inside the finite analysis window of the FIR filter. This approach is susceptible to high-frequency noise; therefore the filter is augmented with a recursive first-order low-pass smoother. However, by reducing the order of the fitted polynomial it is possible to naturally incorporate low-pass behaviour within a regression framework without augmentation. Evaluating the fitted polynomial at a future sample (i.e. predictive) or a past sample (i.e. 'retrodictive') yields a phase lead or a phase lag, respectively. The loss of gain control and phase linearity in the predictive case is usually regarded as an undesirable characteristic; however, this is not the case when the filter is used as a phase lead compensator in a control system.

With finite impulse responses, the polynomial filters described so far all have "finite memories" so that the desired response of the filter is 'incised' with the undesirable response of the finite rectangular window in the frequency domain, i.e. the so-called Dirichlet kernel. Using a "fading memory" so that past samples are gradually forgotten [24], rather than abruptly truncated, 'softens' the effect of the window and yields a 'smoother' frequency response. Fading memory is introduced here using weighted least-squares regression, with a decaying exponential used as the weighting function. This approach naturally gives rise to recursive filter implementations with an infinite impulse response. To maintain a satisfactory transient response in control applications, the (real) pole of the weighing function should obviously be inside, but not too close, to the unit circle. Unlike the recursive FIR filter used in [23], this approach eliminates the possibility of filter instability due to imperfect pole-zero cancellation on the unit circle caused by finite precision and rounding errors. Being a recursive (and possibly parallel) implementation, the proposed approach is also potentially more efficient than non-recursive FIR filters.

Zero-latency fading-memory polynomial filters have been re-discovered and re-derived on numerous occasions in various contexts [24]-[27]. In most treatments, the interpolated polynomial is differentiated to yield estimates of state derivatives. Yet another alternative derivation is given in the next section. The basic approach, however, is extended here to include predictive and retrodictive forms for sinusoidal as well as polynomial basis functions.

4.

Posing the filter design process as a weighted least-squares-regression problem is intended to provide an intuitive framework for designing and tuning compensators of any order with arbitrary phase lead/lag and magnitude pass/stop properties. Specification of the weighting function fixes the location of the filter poles. The least-squares fitting procedure then places the zeros to yield unity gain and the desired group-delay at the design frequencies. Alternative least-squares zero-placement procedures have been used previously to design digital PID controllers [28] and digital compensators [29]. The procedure described here is different to those procedures in several respects; for instance here, the designer is also free to choose the pole location. The pole location determines the decay rate of the transient response and the shape of the frequency response away from the DC region.

Use of polynomial basis functions allows the low-frequency response in the near-DC region to be manipulated precisely. In cases where there are oscillatory modes in the plant (i.e. poles near the unit circle) or where sinusoidal reference or disturbance inputs of known frequency are expected, the ability to specify the phase and magnitude of the compensator far from DC is desirable. For this purpose, the polynomial basis functions are replaced by complex sinusoidal basis functions to yield IIR frequency-sampling filters. Similar FIR schemes have been described previously in the literature [23], [30]; however not in the context of compensator design. FIR implementations again yield "finite memory" filters; recursive FIR implementations yield "sliding" frequency analysers with a "finite memory"; whereas recursive IIR implementations with an "expanding memory" yield the well-known Goertzel filter [31]. Modification of the recursive methods using a weighting function, to yield an implementation with a "fading memory", produces filters that are better suited to control applications as they may be easily tuned to attain appropriate frequency-domain *and* time-domain properties for the desired transient and steady-state behaviour. Least-squares regression ensures that the phase and magnitude requirements are satisfied *exactly* at a nominated set of design frequencies by optimally placing the filter zeros in the $z$ plane. Specification of the weighting-function decay-rate again determines the radial position of the filter poles and the behaviour of the frequency response away from the design frequencies; while the frequency of the each sinusoidal basis function determines their angular positions.

The proposed polynomial and sinusoidal filters are useful as frequency compensation elements, applied to the error signal, inside a controller with one degree-of-freedom (1-DOF). In this configuration, where the plant is discretized using a zero-order hold to yield the pulse transfer function $G_p(z)$, the error transfer function $G_e(z)$ and controller gain $K_e$ in the forward path are tuned to yield satisfactory reference tracking and disturbance rejection behaviour with sufficient stability margins, using either a polynomial or a sinusoidal filter (depending on the nature of the inputs).

In a 2-DOF configuration (see Fig. 1) $G_e(z)$ and $K_e$ are primarily tuned for disturbance rejection and stability; a different filter $G_r(z)$ is then used to "shape" the reference input [32]-[34], with an optional gain factor $K_r$ to help remove offset errors in the absence of an integrator. For instance, a low-pass $G_r(z)$ filter could be used to excise high-frequency content from a step input that would otherwise excite complex poles near the unit circle in the closed-loop system. High-order reference-input filters may be useful in "fly-by-wire"-type control systems, where the plant is required to follow irregular and rapidly changing inputs from an operator, that do not have simple low-order forms.

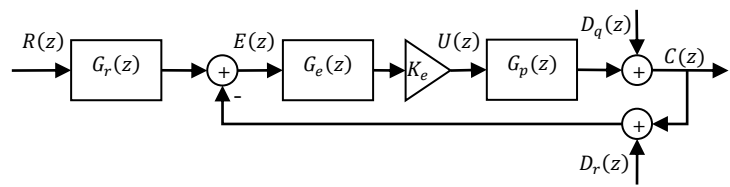

Fig. 1. The assumed 2-DOF controller and system structure. $R(z)$, $C(z)$, $E(z)$, $U(z)$, $D_q(z)$ and $D_r(z)$ are the $Z$ transforms of the reference input $r(n)$, system output $c(n)$, error signal $e(n)$, control signal $u(n)$, plant-disturbance input $d_q(n)$ and sensor-noise input $d_r(n)$, respectively.

## 3. Filter Design

*3.1 Polynomial Filters*

A given digitized signal $x(n)$, is represented over a recent time interval using a $(K+1)$th-order general-linear-model (GLM)

$$x(n-m) = \sum_{k=0}^{K} \beta_k \psi_k(m) + \epsilon \qquad (1)$$

where $\beta$ are the linear coefficients, $\psi$ are the polynomial basis function components $\psi_k(m) = m^k$, $\epsilon$ is a Gaussian noise term with a zero mean and an unknown variance of $\sigma_\epsilon^2$ i.e. $\epsilon \sim \mathcal{N}(0, \sigma_\epsilon^2)$. Recent samples within the finite 'analysis' window of length $M$, are indexed using $m = 0 \dots M-1$, where $m = 0$ corresponds to the most recent sample. The input signal $x(n)$, may be either $r(n)$ or $e(n)$ in Fig. 1, or perhaps $c(n)$ if a compensation element $G_c(z)$ is also used in the feedback path.

The maximum likelihood estimator (MLE) of the linear coefficient vector $\widehat{\boldsymbol{\beta}} = [\hat{\beta}_0, \dots \hat{\beta}_k, \dots \hat{\beta}_K]^T$, where the T superscript is the transpose operator, is found in the usual way, by minimizing the sum of squared errors (SSE) using

$$\widehat{\boldsymbol{\beta}} = (\boldsymbol{\Psi}^T \boldsymbol{\Psi})^{-1} \boldsymbol{\Psi}^T \boldsymbol{x} \qquad (2)$$

where $\boldsymbol{x} = [x(n), \dots x(n-m), \dots x(n-M+1)]^T$

and $\boldsymbol{\Psi}$ is an $M$ by $K+1$ matrix with the element in the $m$th row and $k$th column equal to $\psi_k(m)$. With the MLE coefficients determined via this 'analysis' operation, the estimate of the (noise-free) signal is then reconstructed at an arbitrary point in the future $\acute{m} < 0$, present $\acute{m} = 0$, or past $\acute{m} > 0$, using the 'synthesis' operation

$$\hat{x}(n-\acute{m}) = \sum_{k=0}^{K} \hat{\beta}_k \psi_k(\acute{m}) \qquad (3)$$

or

$$\hat{x}(n-\acute{m}) = \boldsymbol{\psi}\widehat{\boldsymbol{\beta}} \qquad (4)$$

where $\boldsymbol{\psi} = [\psi_0(\acute{m}), \dots \psi_k(\acute{m}), \dots \psi_K(\acute{m})]$, i.e. a 1 by $K+1$ row vector. Note that in general, $\acute{m}$ need not be a round number – using a non-integer delay ($\acute{m} > 0$) results in interpolation if $M = K+1$. The synthesis and analysis operations are combined by substituting (2) into (4) and simplifying, yielding the convolution

$$y(n) = \hat{x}(n-\acute{m}) = \sum_{m=0}^{M-1} h(m)x(n-m) \qquad (5)$$

where $h(m)$ are the coefficients of an FIR filter and $y(n)$ is the output of the filter. The quality of the estimate deteriorates as $\acute{m}$ moves away from the centre of the analysis window i.e. where $\acute{m} = (M-1)/2$ (for a linear-phase filter) and improves as the size of the analysis window increases. Gross errors are expected when the assumed linear model is incorrect. In the frequency domain, estimation errors manifest themselves in the form of deviations away from the desired phase and gain response.

Instead of using a uniform weighting over the analysis window to estimate the model parameters, a non uniform weighting function is now applied to 'de-emphasize' older samples using

$$\widehat{\boldsymbol{\beta}} = (\boldsymbol{\Psi}^T \boldsymbol{W} \boldsymbol{\Psi})^{-1} \boldsymbol{\Psi}^T \boldsymbol{W} \boldsymbol{x} \qquad (6)$$

where $\boldsymbol{W}$ is a square $M$ by $M$ matrix of zeros with the weighting vector $\boldsymbol{w} = [w(0), \dots w(m), \dots w(M-1)]$ along its diagonal. Using $w(m) = e^{\sigma m}$ with $\sigma < 0$, is convenient because it has the desired 'aging' effect, but more importantly, it gives rise to simple $\mathcal{Z}$ transforms.

In the treatment that follows, the 'analysis' operation in (6) is factored into consecutive 'projection' $\mathcal{P}$, and 'orthonormalization' $\mathcal{O}^{-1}$, operators; where $\mathcal{P} = \boldsymbol{\Psi}^T \boldsymbol{W}$ is a $K+1$ by $M$ matrix and $\mathcal{O} = \boldsymbol{\Psi}^T \boldsymbol{W} \boldsymbol{\Psi}$ is a square $K+1$ by $K+1$ matrix. The need for orthonormalization may be avoided if a similarity transform is first applied to create an orthonormal basis-set using the Gram-Schmidt procedure.

If $M$ and $\sigma$ are large, so that $w(m)$ effectively decays to zero at the end of the analysis window, then the FIR and IIR filters have the same impulse and frequency responses, although they will have a very different pole/zero structures.

To derive IIR filters, let $M$ be infinite. The $\mathcal{O}$ matrix then contains infinite summations as its elements with

$$o_{k_2,k_1} = \sum_{m=0}^{\infty} \psi_{k_2}(m) e^{\sigma m} \psi_{k_1}(m) \qquad (7)$$

where $o_{k_2,k_1}$ is the element at the $k_2$th row and $k_1$th column in $\mathcal{O}$, which can conveniently be evaluated in the $z$ domain for polynomial components $\psi_k = m^k$, if $\sigma < 0$, using

$$o_{k_2,k_1} = \mathcal{Z}\{e^{\sigma m} m^{k_1+k_2}\}|_{z=1}. \qquad (8)$$

The projection operator $\mathcal{P}$, is handled in a similar way using

$$p_k(z) = \mathcal{Z}\{e^{\sigma m} m^k\} = \frac{B_k(z)}{A_k(z)} \qquad (9)$$

where $p_k(z)$ is the $k$th element in the vector $\boldsymbol{p}(z)$ and where $B(z)$ and $A(z)$ are polynomials in $z$, thus

$$\boldsymbol{p}(z) = \left[\frac{B_0(z)}{A_0(z)}, \quad \ldots \quad \frac{B_k(z)}{A_k(z)}, \quad \ldots \quad \frac{B_K(z)}{A_K(z)}\right]^T. \qquad (10)$$

Taking the inverse $\mathcal{Z}$ transform of $\boldsymbol{p}(z)$ yields a $K+1$ by 1 column vector containing $K+1$ difference equations in the time domain as elements; thus the infinite summations associated with the projection operation may now be computed recursively using a filter bank, in parallel if desired. The inverse of $\mathcal{O}$ 'mixes' the filter bank outputs to yield the model coefficient vector $\widehat{\boldsymbol{\beta}}$, which completes the analysis operation. This procedure assumes that the model order and the 'memory parameter' $\sigma$ have been selected in a way that ensures $\mathcal{O}$ is a non-singular matrix, which may not be the case if $K$ is large and $\sigma$ is too close to zero, for a very gradual weighting function decay.

Multiplication by the synthesis operator $\boldsymbol{\psi}$ then finalizes the process to yield the filter output $y(n)$. Derivatives of the fitted polynomial may be computed prior to the application of the synthesis operator if desired, leading to the expressions in Table III of [25], for $\acute{m} = 0$ and $K = 1$ (first column) or $K = 2$ (second column).

Instead of a filter-bank implementation, which may be a convenient form in a parallel processor, the inverse $\mathcal{Z}$ transform of the following $z$-domain expression may be used to determine the difference equation of an equivalent high-order filter

$$H(z) = \frac{B(z)}{A(z)} = \boldsymbol{\psi} \mathcal{O}^{-1} \boldsymbol{p}(z). \qquad (11)$$

Note that the polynomial filter has been derived as a low-pass filter due to the low-frequency content of the polynomial components. The filter above may be converted to a high-pass filter by subtracting the estimated value $\hat{x}$, from the measured value $x$, at the synthesis sample $n - \acute{m}$, for integer $\acute{m} > 0$.

*3.2 Sinusoidal Filters*

In this case, the following $(2K + 1)$th-order model is used

$$x(n-m) = \sum_{k=-K}^{k=+K} \beta_k \psi_k(m) + \epsilon \qquad (12)$$

where the basis functions $\psi_k(m)$ are now complex sinusoids $\psi_k(m) = e^{i\omega_k m}$. For notational convenience and continuity with the polynomial case, the angular frequencies are assumed here to be uniformly spaced using $\omega_k = 2\pi k/N$ (radians per sample), where $N$ is an arbitrary integer parameter, although this need not be the case in general.

Following the polynomial treatment, $\mathcal{O}$ and $\boldsymbol{p}(z)$ are populated using

$$o_{k_2,k_1} = \sum_{m=0}^{\infty} \psi_{k_2}^*(m) e^{\sigma m} \psi_{k_1}(m) \qquad (13a)$$
$$= \mathcal{Z}\{e^{(\sigma+i\omega_1-i\omega_2)m}\}\big|_{z=1} \qquad (13b)$$

and

$$p_k(z) = \mathcal{Z}\{e^{\sigma m} \psi_k^*(m)\} = \mathcal{Z}\{e^{(\sigma-i\omega)m}\} = \frac{B_k(z)}{A_k(z)} \qquad (14)$$

where the asterisk superscript represents complex conjugation. Note that the sinusoids form an orthonormal basis set over an interval of $N$ samples if $N = 2K + 1$ and if a unity weight is used over this interval (with zero elsewhere).

7.

The vector of model coefficient estimates $\widehat{\boldsymbol{\beta}}$, produced by the analysis operation may now be interpreted as a 'fading-memory spectrum' as it is analogous to the 'finite-memory spectrum' produced by the sliding discrete-Fourier-transform (SDFT) [31].

To fully exploit the flexibility of the sinusoidal basis set, the synthesis operator is constructed by specifying the magnitude scaling factor $c_k$, and phase shift $\varphi_k$, of each component using

$$\boldsymbol{\psi} = [c_K e^{+i\varphi_K}, \quad \ldots \quad c_k e^{+i\varphi_k} \quad \ldots \quad c_0, \quad \ldots \quad c_k e^{-i\varphi_k} \quad \ldots \quad c_K e^{-i\varphi_K}]. \quad (15)$$

When a fixed phase-delay $\acute{m}$, at all design frequencies is desired, for an approximately linear-phase filter when $N \gg 0$ (for closely-spaced frequency bins) and $\acute{m} > 0$, then $\varphi_k = -\omega_k \acute{m}$ may be used in the above. As in the polynomial case, (13)-(15) are then substituted into (11) to determine the difference equation coefficients.

*3.3 Filter Tuning Principles*

For the sinusoidal filters, application of a large delay ($\acute{m} \gg 0$) decreases estimation errors in the time domain thus *increases* the frequency selectivity of the filter, with reduced deviation in the gain in between the design frequencies in the pass band (for $0 \leq |\omega| \leq \omega_K$) and increased attenuation in the stop band (for $\omega_K < |\omega| \leq \pi$); whereas the use of a rapid decay rate ($\sigma \ll 0$), to produce a filter with a short memory, increases the estimation errors in the time domain thus *decreases* frequency selectivity. The specified delay must therefore be 'supported' by a commensurate filter memory if a lag compensator is to be an effective low-pass filter. To fully exploit a very long filter history, a moderately long delay should be used. Using $\acute{m}\sigma \cong -1$ is a useful 'rule of thumb'. Too much frequency discrimination however, produces a lag compensator with a very 'sluggish' transient response, which has a destabilizing effect in a closed loop configuration. An appropriate balance must therefore be found.

Frequency selectivity is not so important in a lead compensator, which is fortunate because this is very difficult to achieve in the predictive case (i.e. with $\acute{m} < 0$) because estimation errors are much larger than in the interpolative case. Using a shorter memory places the filter poles closer to the origin in the $z$ plane and yields a 'smoother' frequency response, which helps to 'spread' the forward phase shift over the entire frequency range in a lead compensator.

The polynomial filters manipulate the near-DC region of the frequency spectrum. This property of polynomial basis functions is exploited to create maximally flat low-pass FIR filters in [14]. The bandwidth of a polynomial lag filter increases (slowly) with the order of the polynomial model. The frequency selectivity, i.e. the high-frequency attenuation and the width of the transition band, is also governed by the $\acute{m}$ and $\sigma$ parameters.

The frequency response and temporal (steady-state and transient) response of the sinusoidal compensator may therefore be understood in terms of just four design parameters ($K$, $N$, $\acute{m}$ and $\sigma$) or three for the polynomial compensators ($K$, $\acute{m}$ and $\sigma$). The above design guidelines are illustrated by example in the next section.

## 4. Simulated Design Examples

*4.1 Method and results*

The plant to be controlled was formed from two cascaded first-order process – one 'fast', with a pole at $s = -2.5$ s$^{-1}$ (an actuator perhaps); and one 'slow', with a pole at $s = -0.3125$ s$^{-1}$ – yielding the over-damped second-order system

$$G_p(s) = \frac{1}{s^2 + 2.813s + 0.7813}. \quad (16)$$

The continuous-time plant was discretized using the sampling period $T = 0.05$ s and a zero-order hold at the output of the controller to yield

$$G_p(z) = \frac{0.001193z + 0.001139}{z^2 - 1.867z + 0.8688} \quad (17)$$

which has a zero at z = -0.9542 and poles at z = 0.9845 and z = 0.8825. The uncompensated plant has good stability margins (see

8.

Fig. 2) but a relatively slow step response (see Fig. 3). The compensator design process therefore aimed to improve reference input tracking and disturbance input rejection. Lead compensation does this by further increasing the stability margins, thus allowing a very large gain to be applied, for a much improved transient response with degraded noise immunity. Lag compensation allows moderate gains to be applied without amplifying medium- to high-frequency noise.

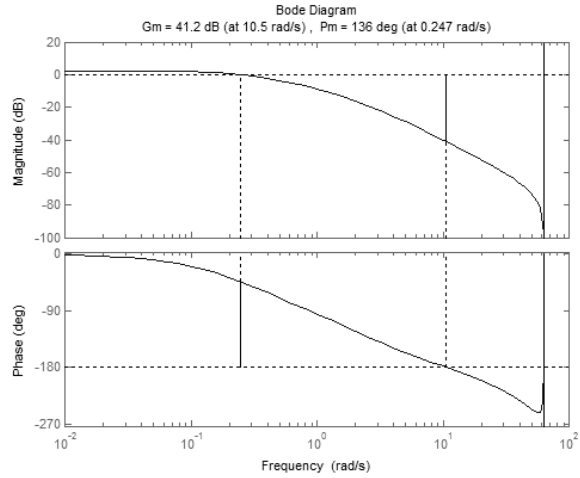

Fig. 2. Bode diagram for the uncompensated plant.

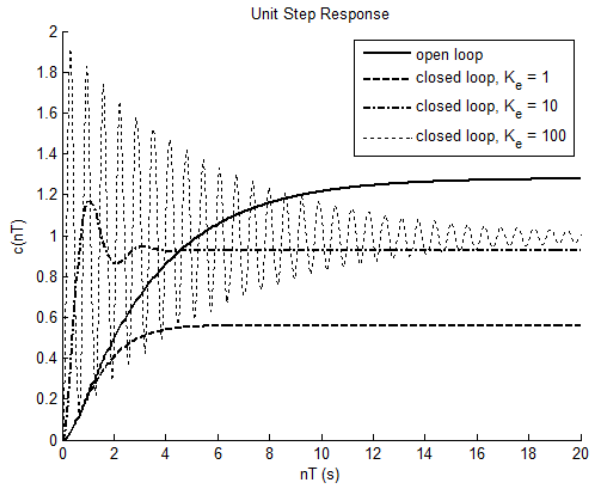

Fig. 3. Unit step response of the uncompensated plant in an open loop configuration and a closed loop configuration with $G_e(z) = 1$ for various gain values $K_e$.

Error-signal lag and lead compensators $G_e(z)$, were designed for this system using both polynomial and sinusoidal filters. To utilize the 2-DOF controller structure, reference shapers $G_r(z)$ were also designed using polynomial lag filters. As the reference shapers are applied outside the feedback loop, they do not affect the stability margins of the system; furthermore, as they are only applied to the reference input, they have no impact on disturbance responses.

The error-signal lag filters were designed to reduce the impact of Gaussian-distributed sensor-noise inputs $d_r(n)$, with $d_r(n) \sim \mathcal{N}(\mu_r, \sigma_r^2)$ to give a uniform power density spectrum over the frequency range. The selected parameters were found to give reasonable medium- to high-frequency noise attenuation without unreasonable destabilization due to an excessively long phase delay.

For the polynomial lag filter, using $K = 1$, $\acute{m} = 2$ and $\sigma = -0.5$ was found to yield satisfactory high-frequency attenuation without introducing a destabilizing delay (see Fig. 4 for the filter response). This yields the following difference-equation

9.

coefficients for the second-order IIR filter (with real poles and zeros):

$b(m) = [0.3225 \quad -0.1677 \quad 0]$

$a(m) = [1 \quad -1.2131 \quad 0.3679]$.

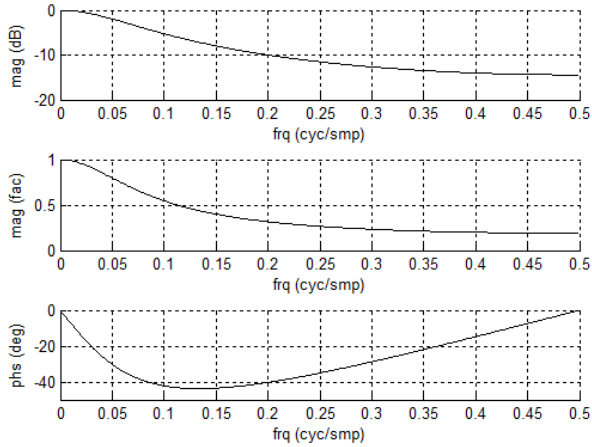

Fig. 4. Frequency response of the polynomial lag filter. Frequency axes in cycles per sample. Upper subplot: magnitude in dB, middle subplot: linear magnitude, lower subplot: phase in degrees.

A second-order sinusoidal lag filter was designed using $K = 1$ and $N = 2$, to specify DC and Nyquist design frequencies; $\sigma = -0.75$ was also used (see Fig. 5 for the filter response). Zero dB gain at DC and -40 dB gain at Nyquist were specified for good high frequency attenuation without too much disruption to the phase response. This yields the following difference-equation coefficients for the IIR filter (with real poles and zeros):

$b(m) = [0.3923 \quad 0.3846 \quad 0]$

$a(m) = [1 \quad 0 \quad -0.2231]$.

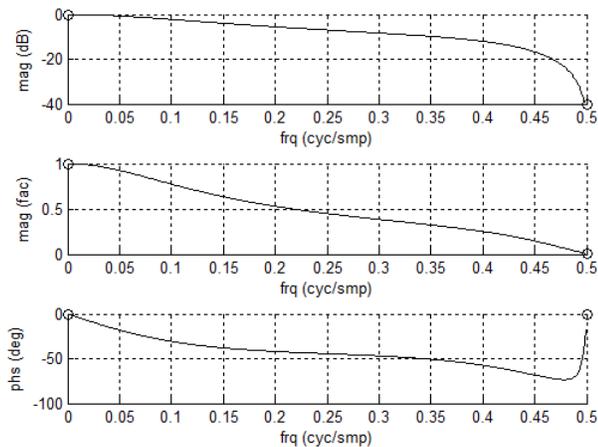

Fig. 5. Frequency response of the sinusoidal lag filter. Specified magnitude and phase values at the design frequencies are shown using circles. See Fig. 4 for subplot description.

The polynomial lead filter was designed using $K = 2$, $\acute{m} = -4$ and $\sigma = -1.5$ to yield a maximum phase lead of approximately 86 degrees (see Fig. 6 for the filter response). These parameters yield flat magnitude and an almost linear-phase response in the near-DC region. The following difference-equation coefficients for a third-order IIR filter result:

$b(m) = [9.1689 \quad -15.1207 \quad 6.4206 \quad 0]$

$a(m) = [1 \quad -0.6694 \quad 0.1494 \quad -0.0111]$.

This filter has real repeated poles, a zero at the origin and a complex conjugate pair of zeros. The locations of the complex zeros tend to stay near $z = 1$ for the polynomial filter. It is easier to shift them to an arbitrary location using the sinusoidal filter; however for his plant, low-frequency zeros are ideal.

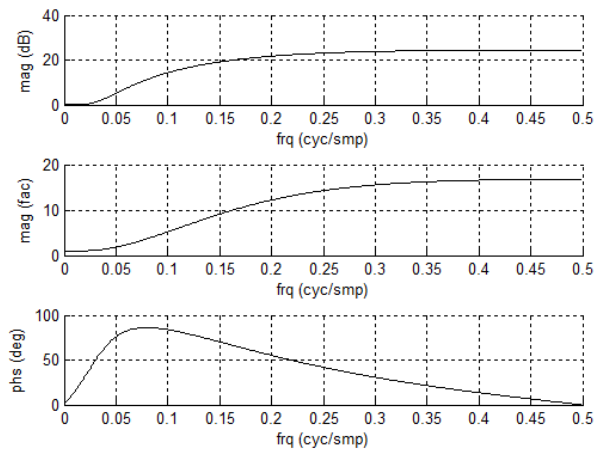

Fig. 6. Frequency response of the polynomial lead filter. See Fig. 4 for subplot description.

The sinusoidal lead filter was designed using $K = 1$ and $N = 16$, to specify design frequencies at DC and 1/16 cycles per sample. At these frequencies the specified gains were -20 dB and 0 dB, respectively; while the specified phase leads were 0 and 90 degrees, respectively. A value of $\sigma = -1$ was also used (see Fig. 7 for the filter response). The following difference-equation coefficients for a third-order IIR filter result:

$b(m) = [2.2228 \quad -3.9018 \quad 1.7078 \quad 0]$

$a(m) = [1 \quad -1.0476 \quad 0.3854 \quad -0.0498]$.

This filter has one real pole and a complex conjugate pair of poles – all with the same radius in the $z$ plane. It also has two real zeros near $z = 1$ and one zero at the $z$-plane origin.

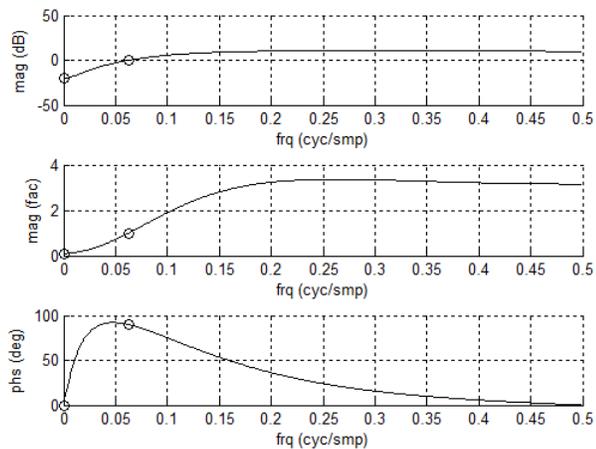

Fig. 7. Frequency response of the sinusoidal lead filter. See Fig. 5 for subplot description.

11.

After the initial loop-shaping filters were designed, the stability margins were then all but 'consumed' by applying a controller gain factor of $K_e$ to reduce steady-state reference- tracking errors and the influence of disturbance inputs. In all cases, $K_e$ was increased until the (linear) gain margin (GM) was at least 2 and the delay margin (DM) was at least 1 sample. The delay margin is computed from the sampling period ($T$, in seconds), phase margin (PM, in degrees) and the gain cross-over frequency ($\omega_{gxo}$, in radians per second) using

$$DM = \left(\frac{\pi}{180}PM\right)/(T\omega_{gxo}). \qquad (18)$$

Linear gains ($K_e$) of 12.5, 20.0, 40.0 and 100.0 were applied with the polynomial lag, sinusoidal lag, polynomial lead and sinusoidal lead filters, respectively.

After the (inner) error filters ($G_e$) were finalized, the (outer) reference filters ($G_r$) were designed to 'sculpt' the response of the closed-loop system to a step reference input. This involved low-pass filtration to varying extents. A polynomial filter with $\acute{m} = 0$, for nominal lag/lead 'neutrality', was used in all cases. The polynomial and sinusoidal lag filters both required reference filters with $\sigma = -0.0125$ for very long memories, to remove high-frequency ripple in the response caused by the diminished delay margins; and $K = 1$ to give a slightly under-damped response. A reference filter with $K = 0$ and $\sigma = -0.250$ was used with the polynomial lead filter; while $K = 1$ and $\sigma = -0.125$ was used with the sinusoidal lead filter.

To assist with compensator tuning, the behavior of the controllers was also examined via discrete-time simulation, using a unit-step reference input $r(n)$, a sensor-noise input $d_r(n)$ with $\mu_r = 0$ and $\sigma_r^2 = 10.0^{-4}$, and various low-frequency sinusoidal plant-disturbance inputs $d_q(n) = \sin(\omega_q n + \varphi_q)$ with $\omega_q \leq \pi/32$ (i.e. an upper limit of 1/64 cycles per sample) and $\varphi_q$ randomly drawn from a uniform distribution over the interval $[0,2\pi]$. Note that the disturbance was added after the plant, as shown if Fig. 1. This forces the controller to manipulate the plant so that the plant output is nearly 180 degrees out of phase with the disturbance, so that the output and the input interfere destructively. Adding the disturbance before the plant is a somewhat easier problem for this system, because the plant naturally attenuates much of the disturbance [9].

*4.2 Discussion*

Given the intrinsic low-pass nature of the plant, there is no real need for a lag compensator to assist with sensor noise reduction. The lag compensators do not improve the response to a reference step input relative to a simple gain-only controller with $K_e = 10$ (compare Fig. 8 with Fig. 3); although, the sinusoidal lag compensator does help to attenuate the sinusoidal disturbance input. Simulations indicated that using a gain-only controller with $G_e = 1$ and $K_e = 10$ reduces the amplitude of the sinusoidal disturbance (with $\omega_q = \pi/32$) to 0.74 at steady-state; using the polynomial lag compensator reduces the amplitude to 0.69; while the sinusoidal lag compensator reduces the amplitude to 0.38. The difference in the lag compensator behavior can be explained be examining the gain responses of the filters in the top subplot of Figs. 4 & 5 – The gain of the sinusoidal filter rolls off more slowly than the polynomial filter in the near-DC region, therefore it is better able to deal with the disturbance.

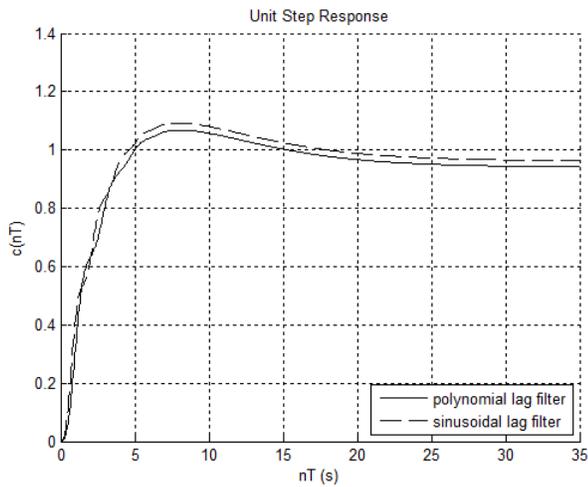

Fig. 8. Response of the closed-loop system with lag compensation for a unit step reference input.

Both lead compensators significantly improve the response to a reference step input relative to a simple gain-only controller with $K_e = 10$ (compare Fig. 9 with Fig. 3). They also result in better disturbance rejection. For the disturbance examined above (with $\omega_q = \pi/32$), the polynomial and sinusoidal lead compensators reduce the amplitude to 0.16 and 0.29, respectively; however, simulations indicate that this is achieved at the expense of sensor noise amplification in both cases. The phase lead profile produced by both compensators is similar; although this is achieved using different gain profiles (compare Fig. 6 with Fig. 7).

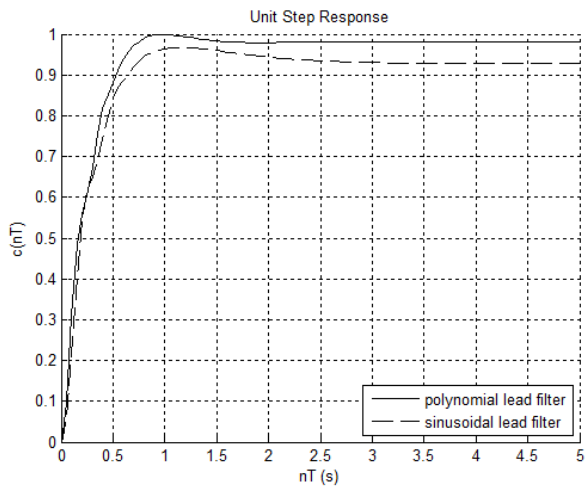

Fig. 9. Response of the closed-loop system with lead compensation for a unit step reference input.

The primary purpose of these simulations is to illustrate the operation of the proposed loop-shaping filters. The filters will of course behave differently in different contexts, depending on whether the system is noise limited (i.e. requires lag compensation) or stability limited (i.e. requires lead compensation). The plant chosen here does not really fall into either category. In general, the polynomial filters are better for low-frequency shaping, whereas the sinusoidal filters are better for shaping the response at arbitrary frequencies.

## 5. Real Design Examples

*5.1 Method and Results*

Polynomial lead and lag compensators were designed and implemented in software running on a personal computer which was used to control a hobby-grade electric motor. In this application, the plant input is the electrical potential (V) and the output is rotational speed, in units of revolutions per second (rps). A disk with five black and five white sectors was attached to the drive shaft of the motor and a photoelectric sensor was placed in front of the disk's face to estimate the rotation rate. The sensor outputs a 4 V signal when a white sector passes by. The analogue pulse-train which is output by the sensor was digitized using an analogue-to-digital (A2D) converter at a rate of 40 kHz and processed using a low-level measurement thread coded using the C programming language. The measurement thread determined the instantaneous inter-pulse period using a simple threshold crossing algorithm then wrote the value to a shared buffer on the arrival of very new pulse.

The buffer was read asynchronously by a high-level control and user-interface thread, coded using the C# programming language. Periodic read events were triggered by a software timer at a rate of 20 Hz. The contents of the buffer were flushed and averaged on each read event. A lock mechanism was used to coordinate buffer access. The control thread then closed the loop by determining an appropriate input voltage for the motor. The control action was sent immediately to the motor via a digital-to-analogue (D2A) and held constant for one sample period. As control-action computation time was negligible, relative to the control loop period ($T$) of 0.05 s, there was little to be gained by waiting for the next control cycle to apply the action. Thus a fast response, possibly with some random variation, was favoured over a slower (delayed) deterministic response [21].

The C# software layer provides a graphical interface and allows the user to select, configure and visualize the operation of a variety of different control algorithms. Digital PID and a digital LSS algorithm were selected for comparison in this work because they are of similar complexity and capability, thus they are all likely to be used in the same sorts of applications. The polynomial lag and lead CoMPensators were used in variants of what will be referred to as the 'CMP' controller.

The PID algorithm was implemented using the conventional "positional form", with all terms in the forward path. The D term was implemented using Euler's backward difference method, $G_d(z) = K_d \frac{(z-1)}{Tz}$; forward differences were used for the I term, $G_i(z) = K_i \frac{Tz}{(z-1)}$ [13].

The LSS algorithm allows closed-loop poles to be arbitrarily placed using internal state feedback while steady-state errors are eliminated using output feedback and an integrator. A full-order current Luenberger observer was used to estimate the internal states [13].

To allow a fair comparison between the algorithms, the lag and lead compensators were combined with a parallel integrator to yield CMP controllers with PI- and PID-type characteristics, respectively.

Three different scenarios were considered –
- *Baseline* scenario;
- *Noise* scenario, where the logic to average the pulse intervals was deactivated, so that the motor speed was inferred using the most recent pulse only; and
- *Delay* scenario where the plant input and plant output were passed through a two-sample delay-line to simulate controller-plant communication delays.

An attempt was made to tune all controllers for reasonable performance in all scenarios. However, to better illustrate the operation of the proposed lag and lead controllers, two CMP implementations were used: one using a polynomial lag filter, for the noise scenario; the other using a polynomial lead filter, for the delay scenario. For a fair comparison, two PID tunings were also used – with and without the D component, for the delay and noise scenarios respectively.

14.

The motor input was restricted to the 0 V to 5 V range; however, the motor reaches a near-maximum speed of approximately 200 rps at around 3 V and the speed only increases slightly as the voltage is further increased. There is also some 'stickiness' at low voltages, as the motor only begins to turn when the initial voltage is above 2 V. It then stops turning when the voltage falls below 1 V.

Reference input step responses for the various controllers are shown in Fig. 10 to Fig. 14. In these figures, a sliding time window of 10 seconds is shown ($x$ axis). The solid black line is the reference input $r(nT)$, and the dashed gray line is the measured plant output $c(nT)$, in revolutions per second, ranging from 0 to 250 rps (scale shown on right-hand $y$ axis). The solid gray line is the plant input voltage, or control signal $u(nT)$, ranging from 0 to 5 V (scale shown on the left-hand $y$ axis). In automatic (closed-loop) mode, the user is able to adjust the reference input $r$, using a slider control in the graphical user interface to give sudden or gradual changes. Input steps of $\pm 50$ rps are shown in the figures. In manual (open-loop) mode, the user adjusts the plant input $u$ directly.

Using the manual mode of operation, the motor input was abruptly increased from 0 V to 3 V to create a step function input for system identification purposes [35]. The least-squares time-domain method described in [13] was then used. A second-order model with two-poles and no zeros fitted the data well. With $T = 0.05$ s, the pole locations were estimated to be at 0.9658 and 0.2717, yielding the following discrete-time model of the plant:

$$G_p(z) = \frac{1.7263}{z^2 - 1.2375z + 0.2624}. \qquad (19)$$

The LSS controller was designed to yield (three) repeated closed-loop poles at $z = 0.75$. Moving the poles closer to the origin for a faster response resulted in excessive noise amplification thus a rapidly fluctuating control signal and an excessively large control signal for step inputs, leading to actuator 'saturation' at 5 V then overshoot due to integrator 'windup'. Logic to handle integrator windup was included in the C# layer, although it was not activated in this study. The observer was designed with repeated poles at $z = 0.25$. After converting (19) into an equivalent state-space representation in controllable canonical form [13], these poles yielded a LSS controller with observer gains

$K_{\text{obs}} = [0.4413 \quad 0.4272]$

state feedback gains

$K_{\text{fbk}} = [0.1595 \quad -0.0125]$

and an integrator gain of

$K_{\text{int}} = 0.0091$.

See Fig. 10 for a plot of the closed-loop system response.

15.

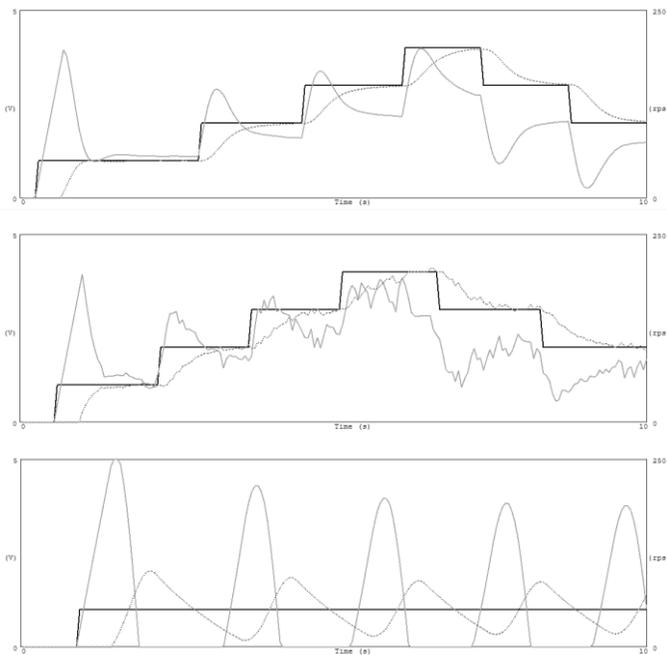

Fig. 10. Response of the LSS motor controller. In the baseline (top), noise (middle), and delay (bottom), scenarios. See text for subplot description.

The PID controller was initially tuned using $K_p = 0.05$, $K_i = 0.05$ and $K_d = 0.00$, to yield a PI controller, for reasonable performance in the baseline and noise scenarios. Relatively low P & I gains and a zero D gain were required to reduce noise amplification. See Fig. 11 for a plot of the closed-loop system response.

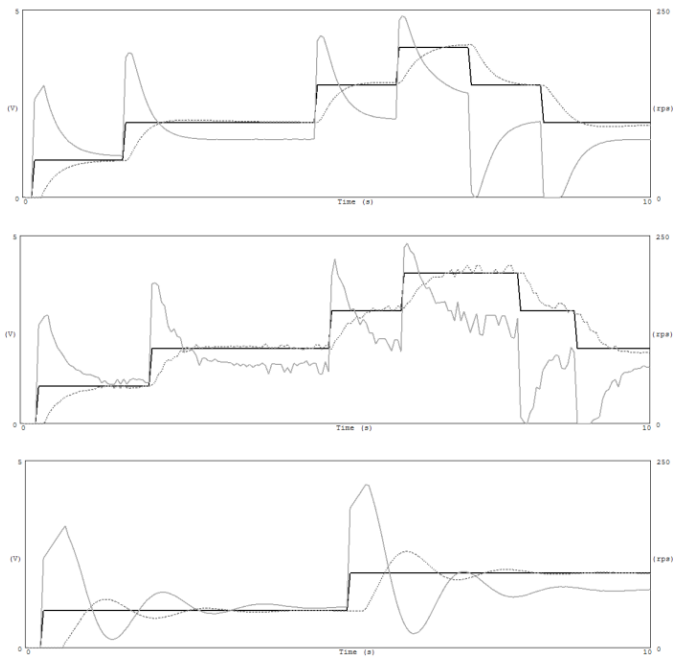

Fig. 11. Response of the PI motor controller. In the baseline (top), noise (middle), and delay (bottom), scenarios. See text for subplot description.

The PID controller was subsequently tuned using $K_p = 0.05$, $K_i = 0.05$ and $K_d = 0.005$ for reasonable performance in the delay scenario. A small amount of derivative control dramatically decreased the settling time in the delay scenario but degraded performance in the other scenarios. See Fig. 12 for a plot of the closed-loop system response.

16.

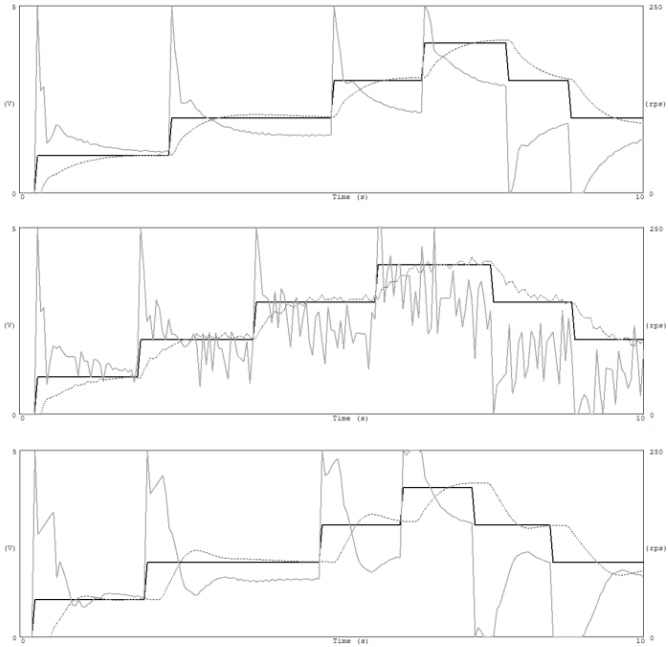

Fig. 12. Response of the PID motor controller. In the baseline (top), noise (middle), and delay (bottom), scenarios. See text for subplot description.

The polynomial lag compensator was tuned to reduce the effects of pulse measurement error in the noise scenario using the $G_e$ parameters from the simulation section. When combined with an integrator using $K_i = 0.05$ and an error gain of $K_e = 0.05$, this gave a nominal gain margin of 5.6573 at 0.0764 cycles per sample and a nominal delay margin of 7.6275 samples at 0.0210 cycles per sample. See Fig. 13 for a plot of the closed-loop system response.

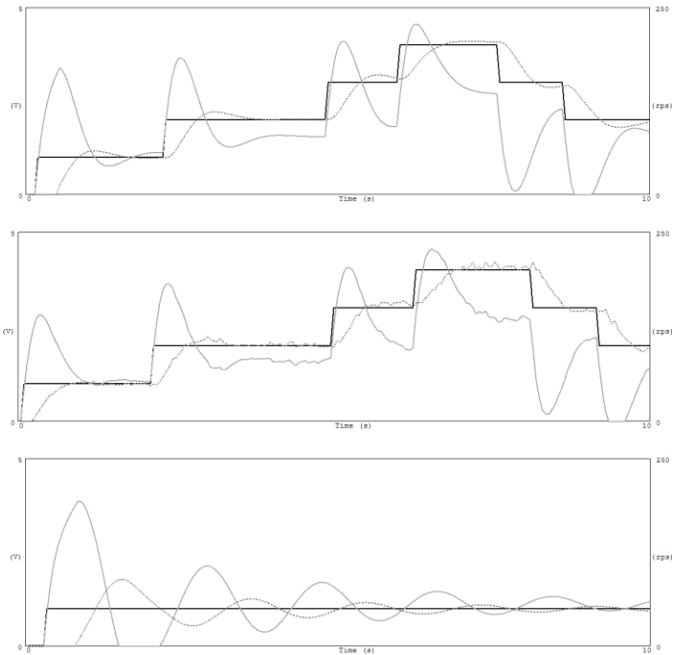

Fig. 13. Response of the CMP motor controller with a polynomial lag filter. In the baseline (top), noise (middle), and delay (bottom), scenarios. See text for subplot description.

17.

The polynomial lead compensator was tuned to improve the stability margins in the delay scenario using $K = 2$, $\acute{m} = -1$ and $\sigma = -1.0$ to create $G_e$. The resulting filter is less severe than the one used in the simulations because $\acute{m}$ is closer to zero. The $G_e$ filter provides a positive phase shift at all frequencies with a maximum phase lead of approximately 30 degrees near 0.1 cycles per sample and a maximum gain of 7.2 dB near 0.25 cycles per sample (compare with the response in Fig. 6). Using the plant model $G_p$ as a guide, and with $G_e$, $K_e$ and $G_i$ combined, this compensator results in a theoretical gain margin of 4.9867 at 0.1751 cycles per sample and a delay margin of 11.0543 samples at 0.0195 cycles per sample for $K_e = 0.05$ and $K_i = 0.05$. As desired, the delay margin of the lead compensator is significantly greater than the delay margin of the lag compensator, for a similar gain margin, which results in greater damping. See Fig. 14 for a plot of the closed-loop system response.

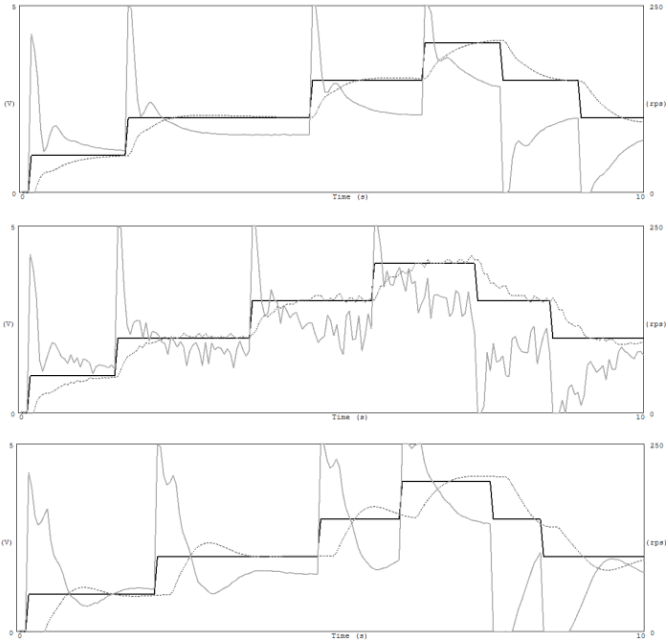

Fig. 14. Response of the CMP motor controller with a polynomial lead filter. In the baseline (top), noise (middle), and delay (bottom), scenarios. See text for subplot description.

The lag and lead compensators both used reference filters with $K = 0$, $\acute{m} = 0$ and $\sigma = -2.0$. These settings result in a moderate $G_r$, with only a slight low-pass effect, primarily due to the short filter memory. Unlike the simulated case, low-pass reference filtering was not required in the real case because the stability margins were greater; therefore, the closed-loop damping for step function inputs was sufficient.

*5.2 Discussion*

Given the flexibility of the three algorithms (PID, LSS and CMP) considered in the motor-control experiment and the lack of any real system requirements, it is difficult to draw meaningful performance conclusions. All algorithms could be tuned to give reasonably similar behavior. However as is already well known and widely appreciated, one obvious difference between the controller types is the ease and speed with which the PID algorithm can be designed and tuned for simple plants. The LSS and CMP controller coefficients were generated using hand-coded Matlab scripts then imported into the C# tool. Once the scripts were written, controller design was a fairly straightforward process. PID does not need this supporting design 'infrastructure', although it may be utilized if available.

Even though the LSS and CMP controllers use a model of the plant, the plant model was not perfect, mainly due to the non-linearities discussed earlier; therefore some empirical tuning was still required. The plant model did however somewhat reduce the time spent tuning. The use of a plant model can be both an advantage and a disadvantage – depending on the quality of the model. The PID controller is unaffected by the model because an empirical tuning approach was used; the LSS controller is strongly affected by the model; whereas the CMP controller allows the model to be used to help the developer understand the relationships between the various conflicting design parameters during the tuning process [9].

If controller development time is the sum of

a) plant modeling;
b) controller coding and implementation; and
c) controller integration, test and tuning activities;

then the production of a good plant model, and the use of a controller design technique that utilizes the derived model, transfers development time from activity c) to activity a). Unlike industrial control problems, for the simple motor control application considered here, there were no safety or productivity penalties associated with online tuning of the closed-loop system. Thus development effort was equally allocated to modeling and tuning activities. This also meant that the appeal of theoretical approaches (such as LSS), or semi-empirical approaches (such as CMP), relative to empirical approaches (such as PID), was somewhat diminished.

Regardless, the simple model and system identification approach appeared to be adequate, as the LSS output more-or-less matched expectations. The predicted closed-loop response of the CMP controller for the model plant also closely matched the actual response for the real plant. Theoretical or predicted step responses, produced via modelling, are plotted in Fig. 15; actual step responses are plotted in the in the top subplot of Fig. 13 and Fig. 14; isolated and exported step responses, from a 'rolling start', are also plotted on the same axes in Fig. 16. According to the system model (see Fig. 15), the lag filter is slightly under-damped; the lead filter has a faster initial rise rate, reduced overshoot but a slightly extended settling time. All of these characteristics are apparent in the response of the real system (see Fig. 16), which indicates that the plant model was reasonable. The theoretical step response and the estimated gain/phase margins could therefore be used to guide and constrain the compensator tuning process.

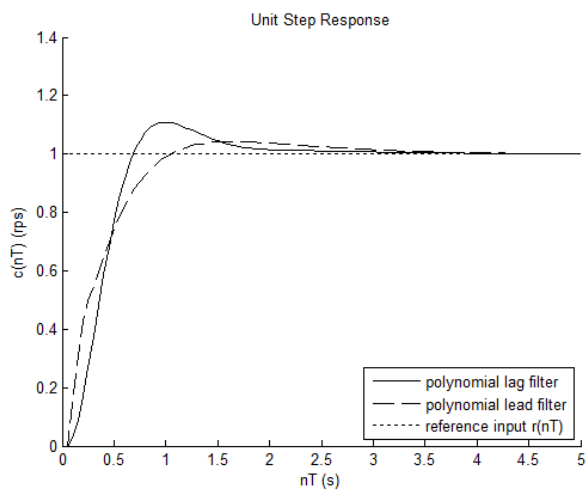

Fig. 15. Predicted response of the CMP controller for two different types of compensator and a unit-step reference input.

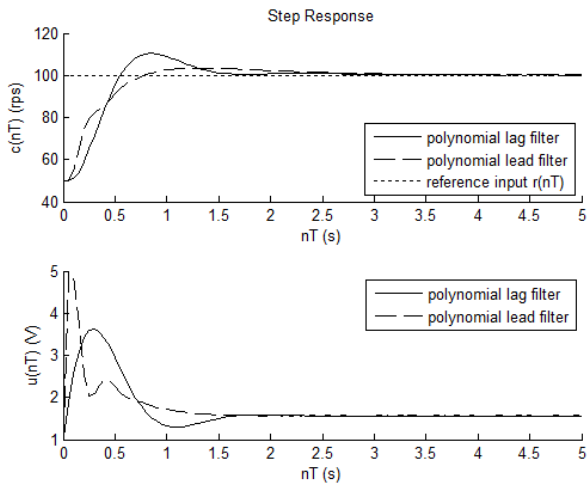

Fig. 16. Actual response of the CMP controller for two different types of compensator and a reference input step of 100 rps.

In addition to the step input, various pulsed and free-form inputs were also tried during system identification. The location of the (dominant) slow pole was reasonably consistent and reproducible, for all input waveforms and on all occasions; the location of the fast pole was somewhat more variable. The step input waveform was mainly used for simplicity and reproducibility. It was also found to be ideal for precise location of the slow pole. More elaborate waveforms may be better for high-order systems [8]; however for various practical reasons, simple steps are still widely used [35].

The time-domain least-squares approach to system identification is perhaps more compatible with LSS controller because the poles and zeros of the plant, thus the form of the recursive observer, are determined directly. The frequency response is then derived by evaluating the pulse transfer function around the unit circle. A frequency-domain system-identification approach, driven by sinusoidal inputs to stimulate critical frequencies, probably would have suited the CMP controller better. Model perfection is not however critical for the successful application of the CMP controller. As a consequence, the designer is able to choose between time spent identifying the system and time spent tuning the compensator.

The CMP controller with the polynomial lead filter gave very similar performance to the PID controller (see Fig. 14 and Fig. 12). Both controllers improved the response in the delay scenario at the expense of the response in the noise scenario. The CMP controller with a polynomial lag filter resulted in slightly better noise suppression than the PI controller; however, the PI step response settled slightly faster in the delay scenario (see Fig. 13 and Fig. 11).

One of advantages in using the lag and lead filters in the CMP controller is the extra loop-shaping flexibility. A single polynomial filter with adjustable lag and lead properties may be used to replace the P and D branches of a PID controller. Furthermore, use of the polynomial filter eliminates the need for customized low-pass derivative filters to reduce noise in the D path. On the one hand, when designing a derivative filter, one of the design issues is the balance between low-frequency phase linearity and high-frequency noise attenuation [16]-[18]; however, in a loop shaping context it is not clear which of these requirements is the more important. On the other hand, when designing a lead filter, the designer only needs to consider gain and phase, using stability margins as a guide.

The LSS controller implementation is slightly more involved than the other controllers. Only first- and second-order variants were coded, using in-line math operations. As a result, no attempt was made to increase the model order to accommodate the system delay, which would presumably have improved the performance in the delay scenario. In the absence of any remedial measures, the LSS controller was only marginally stable in the delay scenario. Had it not been for the actuator saturation which limited the control action, the system probably would have been unstable. The LSS controller performed reasonably well in the

20.

noise scenario and the step response appeared to be critically damped (or at least over-damped) in the baseline scenario, as per the intent of the design (see Fig. 10).

The PID controller performed much better than the LSS controller in the delay scenario, but due to the use of the D term, it performed much worse than the LSS controller in the noise scenario. The non-zero D term was also responsible for the spike (or "kick") in the control signal when the step reference input is first applied (see Fig. 12). Of all the controllers, PID was the easiest controller to tune for a given scenario. However, the CMP controller has many more degrees of tuning freedom, due to the filter design process, which can be both a good and bad feature, depending on the circumstances. The lack of PID flexibility was most evident in the noise scenario, where little could be done in the PI filter to reduce noise amplification, other than setting $K_d$ to zero scenario (see the middle subplot of Fig. 11). Reducing $K_p$ further did help somewhat in this respect but it also slowed the transient response in the baseline scenario.

The logic to average the pulse periods in the C# layer, common to all controllers, had a significant impact on the dynamics of the closed-loop system. This process is part of the controller; however, it appears as part of the plant, because it is required to generate the output for observation. The measurement process is therefore integrated with the plant and included in the system identification process, giving rise to the fast pole of the plant. Smoothing of some kind is essential to reduce noise and to fully utilize all pulse measurements that are collected by the C layer over each period of the C# timer – around 70 pulses per timer tick when the motor is operating near maximum speed. At these speeds, even when the pulse train is digitized at rate of 40 kHz, an error of just one sample in the identification of the pulse edges in the C layer potentially results in a speed error of around 10% in the C# layer. Smoothing the pulse-period measurements hides these errors. Various other smoothing algorithms were considered – for instance, a first-order low-pass IIR filter in the C layer. While this approach allows the behavior of the smoother to be configured, averaging all pulses over a finite time interval was found to give a smoother output and better closed-loop performance in general. As the degree of smoothing decreased, the measurement pole moved to the left along the real $z$ axis. Decreased smoothing increased the stability margins and allowed faster responses to achieved; however, it also increased measurement noise and degraded steady-state performance; as a consequence, derivative and phase-lead filters could not be used effectively in any of the scenarios.

### 6. Conclusion

The polynomial and sinusoidal lag and lead compensators considered in this paper have a number of interesting properties. If a digital controller is required and if frequency-domain design-approach is preferred, then the proposed method is ideal because it allows low-order filters to be quickly synthesized with the requisite magnitude and phase characteristics. The filters may be designed to have arbitrary low-pass, high-pass or band-pass properties, with forward or backward phase shifts over specified frequency bands. Gain and phase requirements are satisfied exactly at specified design frequencies. Designing digital IIR filters using regression analysis allows the gain, phase and frequency properties of a filter/compensator to be understood in the time domain. The polynomial filters assume that the design frequency is zero, thus they give good control over the near-DC frequency region; whereas the sinusoidal filters allow the response at arbitrary frequencies to be specified. The sinusoidal filters result in a design approach that is similar to the "frequency-sampling" DSP filter-design method [10]; however, it is modified here to yield IIR filters rather than FIR filters; furthermore, unlike the "windowing" DSP filter-design method [10], arbitrary tapers are not required to improve the response. In general, other general-purpose optimal DSP design methods are difficult to use in control applications because they do not directly consider requirements such as transient response and stability margins.

When compared with other compensator design techniques, the proposed method:
- Does not suffer from distortion associated with *s*-to-*z* mappings.

21.

- Is more flexible than simpler methods involving closed-form expressions for first- and second-order components.
- Eliminates the need for guesswork when cascading multiple low-order units or when manually assigning single poles or zeros in the *z* plane.

The PID controller, the LSS controller, and the CMP controller involving the polynomial lag and lead filters, could all be tuned for similar performance in the simple simulated and real scenarios considered here. However, the CMP controller has the greatest number of tuning parameters, due to the flexible filter-design process. This feature was found to be most useful in the noise scenario where the lag compensator was able to slightly outperform the PI filter, albeit at the expense of performance in the delay scenario.

In broader control engineering problems, the best solution is determined by many constraints that were not contemplated in this study, it is therefore difficult and unwise to draw definite performance conclusions here. The experiments performed did however reveal the following:

- It was gratifying to use the PID algorithm because a very good controller could be designed and implemented with a minimum of time and effort.
- It was satisfying to use the LSS algorithm because less guess work was required to tune the controller due to the utilization of the plant model; however, a few design iterations were still required.
- It was reassuring to use the CMP algorithm because the plant model and frequency analysis could be used to
    - Justify an initial controller design,
    - Guide the fine-tuning process and
    - Set approximate upper bounds on the controller parameters via the stability margins.

Like PID and LSS, the proposed digital filters may be used to design simple, effective and flexible controllers.

22.